\documentclass[aps,prb,twocolumn,superscriptaddress,showpacs,showkeys]{revtex4}
\usepackage{graphicx}
\usepackage{dcolumn}
\usepackage{bm}

\begin{document}

\preprint{APS/123-QED}

\title{Cyclotron Resonance study of the electron and hole
velocity in graphene monolayers}
\author{R.S. Deacon}
\author{K-C. Chuang}
\author{R. J. Nicholas}

\email{r.nicholas1@physics.ox.ac.uk}\affiliation{Clarendon
Laboratory, Physics Department, Oxford University, Parks Road,
Oxford OX1 3PU, U.K.}

\author{K.S. Novoselov}
\author{A.K. Geim}
\affiliation{Manchester Centre for Mesoscience and Nanotechnology,
University of Manchester, Manchester M19 9PL, U.K.}

\date{\today}
\begin{abstract}

We report studies of cyclotron resonance in monolayer graphene.
Cyclotron resonance is detected using the photoconductive response
of the sample for several different Landau level occupancies. The
experiments measure an electron velocity at the K- (Dirac) point
of $c_{K}^{*}$ = 1.093 x 10$^{6}$ ms$^{-1}$ which is substantially
larger than in thicker graphitic systems. In addition we observe a
significant asymmetry between the electron and hole bands, leading
to a difference in the electron and hole velocities of 5$\%$ by
energies of 125 meV away from the Dirac point.
\end{abstract}

\maketitle

The observation of two dimensional electronic systems in monolayer
graphene \cite{novos04}, where the electrons behave as Dirac
Fermions and show a variety of novel quantum Hall
effects\cite{novos05},\cite{zhang:2005},\cite{zhang:2006}, has led
to an explosion of interest in this system. As well as new basic
science, the exceptionally high electron velocities also mean that
graphene has considerable potential for applications in high speed
electronics\cite{novos07}. The basis for this behaviour is the
nearly linear dispersion of the energy bands close to the K point,
where the dispersion relations cross with the form $E = \pm
c^{*}\hbar k$, where $c^{*}$ is the electron velocity. This has
been predicted for over 50 years \cite{wallace:1947}, but has only
been measured recently for bulk graphite\cite{zhou:2006} and
ultrathin graphite layers \cite{sadowski:2006}, while the first
direct absorption measurements for monolayer graphene have just
been reported\cite{jiang:2007}. We describe here a photoconuctance
study of cyclotron resonance in a monolayer of graphene in which
the application of a magnetic field leads to the formation of
Landau levels given by\cite{mcclure:1957}

\begin{equation}
  E_{N} = \textrm{sgn(N)}\times c^{*} \sqrt{2e\hbar B|N|},
  \label{LL}
  \end{equation}

\noindent where $|N|$ is the Landau quantum index and B is the
magnetic field. This allows us to make a precise measurement of
the electron velocity and to examine deviations from exact linear
behaviour which show that the electron and hole-like parts of the
band structure have significantly different masses and that the
velocity is significantly larger than for thicker graphitic
material.

\begin{figure}
\begin{center}
\includegraphics[width=0.7\linewidth]{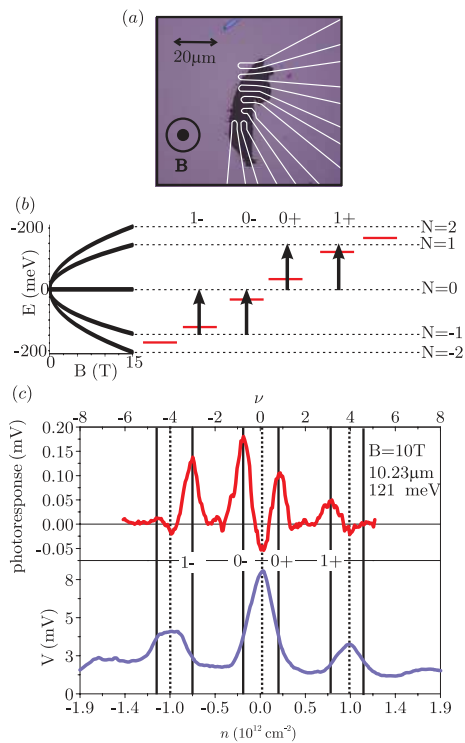}
\caption{\label{PC} (Color online)(\textit{a}) Sample image with
outline of the contacts used in the measurements.(\textit{b}) plot
of the Landau energies as a function of magnetic field for Landau
index $N=-2\ldots 2$. Arrows indicate the resonant transitions
probed in the energy range of the CO$_{2}$ laser. (\textit{c})
Density dependence of the two contact resistive voltage and
photoconductive response of a typical graphene sample for infrared
radiation of 10.23\,$\mu m$ at 10\,T measured with a current of
100nA.}
\end{center}
\end{figure}

The experiment studies the photoconductive response from a
multiply contacted single monolayer sample of graphene, which was
prepared using the techniques which have been described
earlier\cite{novos04,novos05}. The graphene films were deposited
by micromechanical cleavage of graphite with multi-terminal
devices produced by conventional microfabrication, with a typical
sample displayed in figure \ref{PC} (\textit{a}). Shubnikov-de
Haas oscillations were first studied at 1.5K to establish the
carrier densities as a function of gate voltage and to ensure that
the film studied was a single monolayer of graphene, since
bilayers and thicker films are known to have a completely
different dispersion
relation\cite{novosolev:2006},\cite{mccann:2006},\cite{guinea:2006}.

Cyclotron resonance was measured by detecting the modulation of
the conductivity of the samples produced by chopped infrared
radiation from a $CO_{2}$ laser operating between 9.2 and 10.8
$\mu m$. The sample was illuminated normally with unpolarized
light parallel to the magnetic field in the Faraday geometry.
Typical power densities were $\sim$3 x 10$^{4}$ Wm$^{-2}$,
corresponding to a total power incident on the samples studied of
order 5 $\mu$W . The majority of experiments were performed in
two-contact mode with a current of I=100nA since this gave the
best signal to noise ratio, although similar spectra were also
observed in a four contact configuration. Figure \ref{PC}
(\textit{c}) shows the photoconductive signal and the 2 contact
resistance of a graphene layer as a function of carrier density,
$n$, with the sample immersed in liquid helium at 1.5\,K. This
demonstrates that large positive photoconductive signals are
observed at the edges of the conductance peaks, at the points
where the resistivity is changing most rapidly with temperature
and chemical potential. The response is proportional to the energy
absorbed and thus provides an accurate relative measurement of the
absorption coefficient.  At resonance we observe voltage
modulations as high as $3\%$. The peak response is detected when
the Landau level occupancy, $\nu=nh/eB$ is -3.0 (1-), -0.76 (0-),
0.88 (0+) and 3.1 (1+), where 0 corresponds to the Dirac point. A
small negative response is also observed when the Landau levels
are exactly half filled at occupancies of $\nu$= -4, 0, +4. The
two response peaks labelled (1-) and (1+) correspond to hole and
electron-like transitions from the Dirac point (N = 0) to the N =
$\pm$1 Landau levels respectively. The (0-) and (0+) peaks both
correspond to mixtures of the two transitions as the N=0 level is
partially filled with either holes or electrons, but with either
the hole or electron transition transition respectively
predominant as indicated in figure \ref{PC}(\textit{b}). When
$|\nu|>4$ no resonant absorption can occur in this field range and
we only observe some much weaker additional features caused by non
resonant bolometric response from the sample. This is greater at
higher magnetic fields where localisation of the carriers is
increased.

\begin{figure}
\begin{center}
\includegraphics[width=0.7\linewidth]{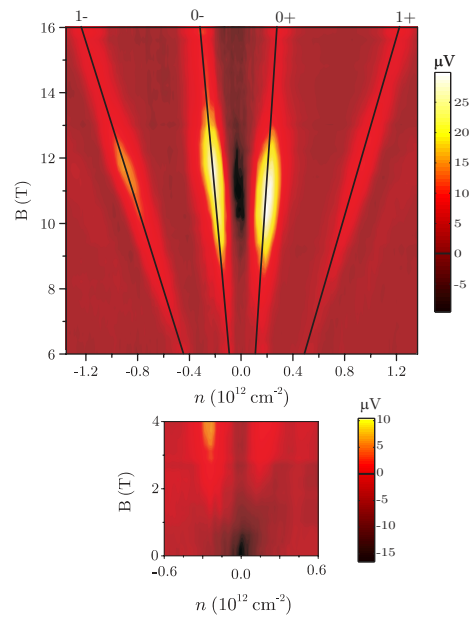}
\caption{\label{CRmap} (Color online) Photoconductive response as
a function of gate voltage and magnetic field for 9.25 $\mu
m$(134meV). The low field section of the map has an enhanced
sensitivity to display the sharp negative resonance at zero
field.}
\end{center}
\end{figure}

In order to detect the resonances we measure carrier density
sweeps at each value of magnetic field, and compile a full map of
the photoresponse as shown in Figure \ref{CRmap} for a wavelength
of 9.25\,$\mu m$. This demonstrates that clear resonances can be
detected for all four occupancies where strong photoresponse is
seen. The immediate conclusion from this plot is that the
resonances all occur in the region of 10T, but that there is a
significant asymmetry between the electron and hole-like
transitions. A further negative photoresponse is observed at low
magnetic fields ($<2$\,T) which we attribute to inter-band photon
absorption processes such as $-(N+1)\rightarrow N$ and
$-N\rightarrow (N+1)$. In order to demonstrate the high field
resonances more clearly and to investigate the magnetic field
dependence of the transition energies we show traces in which the
Landau level occupancy is held constant, by the simultaneous
scanning of the gate voltage and magnetic field in order to follow
the constant occupancy lines as shown in Fig. \ref{CRmap}.

\begin{figure*}[t]
\begin{center}
\includegraphics[width=0.70\linewidth]{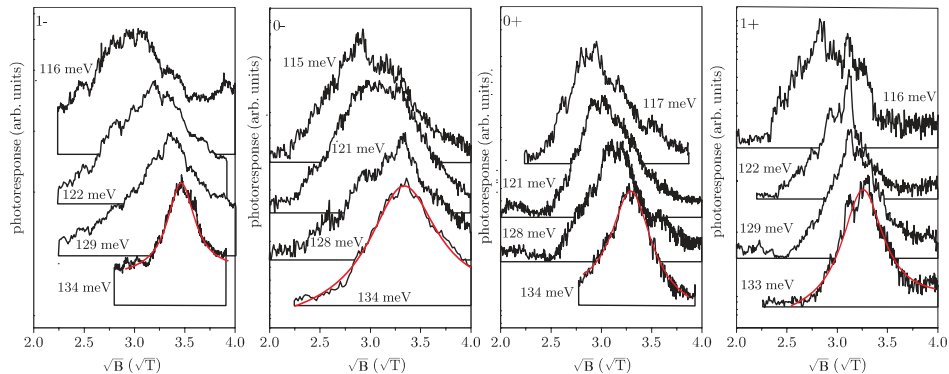}
\caption{\label{CRwaterfall} (Color online) Photoconductive
response as a function of $\sqrt{B}$ with the carrier densities
scanned to keep the occupancies constant at $\nu$= -3.0 (1-),
-0.76 (0-), 0.88 (0+) and 3.1 (1+) for wavelengths from 9.2 to
10.7 $\mu m$. The red lines show fits using Lorenzian lineshapes
combined with a linear background response.}
\end{center}
\end{figure*}

\begin{figure}
\begin{center}
\includegraphics[width=0.8\linewidth]{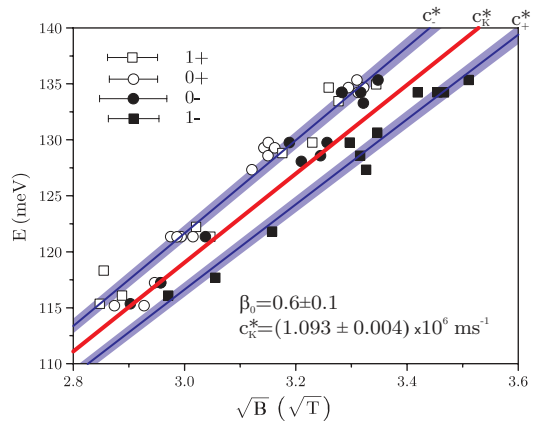}
\caption{\label{resonance} (Color online) Resonance positions for
the four resonances as a function of $\sqrt{B}$, together with a
single fitted value of the electron velocity c$_{K}^{*}$(red
line). The outer lines show fits to equation \ref{CKpm}, with the
shaded bands covering the error limits from c$_{K}$ and
$\beta_{0}$. The individual resonance positions have errors as
shown of $\sim \pm$2\%, corresponding to 0.2 $\Delta\sqrt{B}$,
where $\Delta\sqrt{B}$ is the half width at half maximum
absorption.}
\end{center}
\end{figure}

Sequences of resonances for the electron-like and hole-like
transitions are shown in Fig. \ref{CRwaterfall}. The resonances
are plotted as a function of $\sqrt{B}$ and fitted with
conventional Lorenzian lineshapes with the addition of a linear
correction to account for the increasing bolometric response at
high fields. Some resonances show significant anisotropy and we
therefore quote an error for individual points of $\pm$20\% of the
half width at half maximum. A typical fit is shown for each of the
four resonances. The 0- resonances are particularly broad and
therefore give higher errors. The resonance positions are plotted
as a function of $\sqrt{B}$ in Fig. \ref{resonance}. The resonance
energies are expected to be given by equation \ref{LL}, with a
single value of the electron velocity c$^{*}$. Our results show
clearly that this is not the case. Fitting velocities to each of
the resonances separately gives values of c$^{*}$=(1.117, 1.118,
1.105 and 1.069 $\pm$ 0.004) x 10$^{6}$ ms$^{-1}$ for the 1+, 0+,
0- and 1- resonances respectively. The resonances measured for the
1- and 1+ occupancies show the lowest and highest values for
c$^{*}$ as would be expected if the electron and hole masses are
different, since these correspond to pure hole-like and
electron-like transitions, while the values for (0-) and (0+) are
intermediate between the two extremes. Defining a single Fermi
velocity averaged over the extremal values for electrons and holes
in the region of the Dirac point gives c$^{*}$=1.093 $\pm$ 0.004 x
10$^{6}$ ms$^{-1}$. Interpreting the resonance positions in terms
of the conventional cyclotron effective mass gives m$^{*}$ =
0.009m$_{e}$.

Values reported previously for the Fermi velocity suggest that it
is quite strongly dependent on the number of graphene sheets in
metallic systems.  Angle resolved photoemission on bulk
graphite\cite{zhou:2006} gives 0.91 x 10$^{6}$ ms$^{-1}$, while
the cylotron resonance measurements of Sadowski et
al\cite{sadowski:2006} on thin (3-5) layers of epitaxial graphite
give 1.03 x 10$^{6}$ ms$^{-1}$. A recent report on tunelling
measurements in bilayer graphene\cite{GLi:2007} has found 1.07 x
10$^{6}$ ms$^{-1}$ while the results above and the cyclotron
absorption by Jiang et al\cite{jiang:2007} on monolayer
graphene\cite{zhang:2005} give values of $\simeq$1.1 x 10$^{6}$
ms$^{-1}$. By contrast estimates based on the electronic
properties of semiconducting carbon nanotubes deduce $c_{K}^{*}$ =
0.94 x 10$^{6}$ ms$^{-1}$ corresponding to values of $\gamma_{0}$,
the transfer integral, of order 2.9
eV\cite{milnera:2000,souza:2004}.

Theoretically nearest neighbour tight binding
theory\cite{saito:1998} predicts electron energies in terms of
$\gamma_{0}$ and $s_{0}$, the nearest neighbour overlap integral,
of

\begin{equation}
  E = \frac{\epsilon_{2p} \mp \gamma_{0}\sqrt{\omega (k)}}{1 \mp s_{0}\sqrt{\omega
  (k)}}.
  \label{TB}
  \end{equation}

Setting $\epsilon_{2p}$ = 0 to give a correct description of the
bands close to the K point, and with $\sqrt{\omega (k)} =
\frac{\sqrt{3}}{2}ka_{0}$, where $a_{0}$=0.246 nm is the graphene
lattice parameter,  gives the electron velocity as

\begin{equation}
  c_{\pm}^{*} = c_{K}^{*} \frac{1}{1 \mp
  \frac{s_{0}E}{\gamma_{0}}},
  \label{CKpm}
  \end{equation}

where $c_{K}^{*} =  \frac{\sqrt{3}}{2}
\frac{\gamma_{0}a_{0}}{\hbar}$. Typical values for the parameters
of $\gamma_{0}$=3.03 eV and $s_{0}$=0.129 which have been derived
from first principles calculations \cite{saito:1992} and found to
give good agreement with experiment \cite{saito:1998} give values
for $c_{K}^{*}$ = 0.98 x 10$^{6}$ ms$^{-1}$ but predict only a
very small asymmetry of the velocity of $\pm$ 0.5 $\%$. More
complex calculations such as those including up to third nearest
neighbours\cite{reich:2002} conclude values which lead to even
lower values of $\gamma_{0}$ (2.7 eV) and hence $c^{*}$. This
suggests therefore that the currently accepted values of the
transfer integral are consistent with the graphite results, but
there is a progressive increase in the electron velocity as the
graphite is thinned down to the single monolayer graphene result.
The changes in the transfer integral are probably related to the
screening or changes in the details of the $\pi$ bonds
perpendicular to the graphene surface, which are also responsible
for the band structure at the K point. These bonds are directly
linked to the inter layer coupling of the graphene sheets and to
their coupling to the SiO$_{2}$ insulator, suggesting that this
coupling leads to an enhancement of the electron velocity as has
been suggested recently for carbon nanotubes\cite{li:2006} where
filling of the nanotubes with crystalline material leads to
changes in the transfer integral. Using a value of $c_{K}^{*}$ =
1.093 x 10$^{6}$ ms$^{-1}$ leads to the deduction of a value of
$\gamma_{0}$=3.38 eV.

%

The second conclusion from Figure \ref{resonance} is that the
asymmetry between electron and hole is considerably larger than
that predicted by the simple tight binding theory. We model this
by replacing the overlap integral, $s_{0}$ with an empirical
factor $\beta_{0}$ in equation \ref{CKpm} and re-fitting the data
shown in Figure \ref{resonance} with the modified equation

\begin{equation}
  c_{\pm}^{*} = c_{K}^{*} \frac{1}{1 \mp
  \frac{\beta_{0}E}{\gamma_{0}}}.
  \label{CKpbeta}
  \end{equation}

\noindent The best fits to the data are shown in Fig.
\ref{resonance} with values of $c_{K}^{*}$ = (1.093 $\pm$ 0.004) x
10$^{6}$ ms$^{-1}$ and $\beta_{0}$ = 0.6 $\pm$0.1. These values
give velocities for the electrons and holes of $c_{\pm}^{*}$ of
1.118 and 1.069 x 10$^{6}$ ms$^{-1}$ in the energy range close to
$\pm$ 125 meV. We therefore have clear evidence for the breaking
of particle-antiparticle symmetry in the graphene system at the
level of $\pm$ 2.5 $\%$, approximately five times larger than
expected for simple tight binding theory\cite{saito:1998}. This
may be linked to the intrinsic single particle band structure,
with some indications of this in the comparison of ab initio and
tight binding dispersions\cite{reich:2002}, although these
calculations suggest values of $c_{K}^{*}$ as low as 0.87 x
10$^{6}$ ms$^{-1}$. By contrast the magnitude of the asymmetry is
comparable, but of the opposite sign to that predicted ($\sim \mp$
3 $\%$) using random phase approximation methods which take
account of dynamical screening\cite{miyake:2003}, and which also
predict an overall $\sim 13\%$ enhancement of the velocity. It is
also possible that the gating process itself will lead to some
changes in the $\pi$ bonding due to the changes in surface field
and that this is linked to the velocity enhancement in thinner
layers.

In addition to conventional single particle effects it may also be
possible that many-body corrections could influence the value and
asymmetry of the electron velocity. Kohn's theorem\cite{kohn:1961}
has long been known to exclude the influence of electron-electron
interactions on long wavelength excitations for conventional
parabolic systems. Calculations for graphene\cite{iyengar:2006}
suggest however that although there are several similarities with
the normal electron case, the linear dispersion may lead to finite
Coulomb contributions to the cyclotron resonance transition
energies and that these will be strongly dependent on the level
occupancy, although these are based on perfect particle-hole
symmetry.

The resonance linewidths (half width at half maximum) deduced from
fitting the data in Figure \ref{CRwaterfall} are all in the region
of 0.27 - 0.37 $\sqrt{T}$ (1.5-2.5 Tesla). Using our measured
value of $c_{K}^{*}$ gives an energy broadening $\hbar/\tau$
$\simeq$ 12 meV, corresponding to a simple momentum relaxation
time of $\sim$ 5.5 x 10$^{-14}$s and a mean free path $\lambda$ =
$c^{*}\tau \sim 0.06 \mu$m and a mobility $\mu \sim 1.1
m^{2}Vs^{-1}$. The linewidths are significantly smaller than those
observed by Jiang et al\cite{jiang:2007} which may explain why
these authors did not observe the electron-hole asymmetry.

In conclusion therefore we have measured cyclotron resonance in a
monolayer graphene system, which demonstrates that the electron
velocity is significantly enhanced relative to the value expected
from previous calculations and measurements for thicker graphitic
systems. In addition we have demonstrated a considerable asymmetry
in the carrier velocity for the electron and hole like parts of
the dispersion relation close to the K-point of the Brillouin
zone. These measurements suggest that there are still considerable
uncertainties in understanding the band structure of monolayer
graphene which may lead to significant changes in any theories
\cite{cheianov:2007} based on perfect particle-antiparticle
symmetry.

\section{acknowledgments} Part of this work has been supported by
EuroMagNET under the EU contract RII3-CT-2004-506239 of the 6th
Framework 'Structuring the European Research Area, Research
Infrastructures Action'.


\end{document}